\documentclass[a4paper,11pt]{article}
\pdfoutput=1 

\usepackage{jheppub} 

\usepackage[T1]{fontenc} 

\usepackage{subcaption}
\usepackage{multirow}
\usepackage{soul}

\newcommand{\ttH}{\text{t}\bar{\text{t}}\text{H}}
\newcommand{\ttHbb}{\text{t}\bar{\text{t}}\text{H}(\text{b}\bar{\text{b}})}
\newcommand{\ttbb}{\text{t}\bar{\text{t}}\text{b}\bar{\text{b}}}
\newcommand{\bb}{\text{b}\bar{\text{b}}}
\DeclareMathOperator*{\argmax}{argmax} 

\title{\boldmath Learning to increase matching efficiency in identifying additional b-jets in the $\ttbb$ process}

\author[a]{Cheongjae Jang,}
\author[b]{Sang-Kyun Ko,}
\author[a,b]{Yung-Kyun Noh,}
\author[c]{Jieun Choi,}
\author[c]{Jongwon Lim}
\author[c,1]{and Tae Jeong Kim\note{Corresponding author.}}

\affiliation[a]{A.I. Institute, Hanyang University,\\222, Wangsimni-ro, Seongdong-gu, Seoul, Republic of Korea}
\affiliation[b]{Department of Computer Science, Hanyang University,\\222, Wangsimni-ro, Seongdong-gu, Seoul, Republic of Korea}
\affiliation[c]{Department of Physics, Hanyang University,\\222, Wangsimni-ro, Seongdong-gu, Seoul, Republic of Korea}

\emailAdd{cjjang@hanyang.ac.kr}
\emailAdd{highsg19101@gmail.com}
\emailAdd{nohyung@hanyang.ac.kr}
\emailAdd{ji.eun.choi@cern.ch}
\emailAdd{jongwon.lim@cern.ch}
\emailAdd{taekim@hanyang.ac.kr}

\abstract{The $\ttHbb$ process is an essential channel to reveal the Higgs properties but has an irreducible background from the $\ttbb$ process, which produces a top quark pair in association with a b quark pair. Therefore, understanding the $\ttbb$ process is crucial for improving the sensitivity of a search for the $\ttHbb$ process. To this end, when measuring the differential cross-section of the $\ttbb$ process, we need to distinguish the b-jets originated from top quark decays, and additional b-jets originated from gluon splitting. Since there are no simple identification rules, we adopt deep learning methods to learn from data to identify the additional b-jets from the $\ttbb$ events. Specifically, by exploiting the special structure of the $\ttbb$ event data, we propose several loss functions that can be minimized to directly increase the matching efficiency, the accuracy of identifying additional b-jets. We discuss the difference between our method and another deep learning-based approach based on binary classification \cite{choi2020identification} using synthetic data. We then verify that additional b-jets can be identified more accurately by increasing matching efficiency directly rather than the binary classification accuracy, using simulated $\ttbb$ event data in the lepton+jets channel from pp collision at $\sqrt{s}$ = 13 TeV.}


\begin{document} 
\maketitle
\flushbottom

\section{Introduction}
\label{sec:intro}

Since discovering the Higgs boson at the large hadron collider (LHC) \cite{aad2012observation, chatrchyan2012observation}, its consistency with the standard model has been tested extensively in many different channels. In 2018, there were observations of Higgs boson production in association with a top quark pair ($\ttH$), which is an important channel in revealing the Higgs boson properties \cite{sirunyan2018observation, aaboud2018observation}. As the branching fraction of the Higgs boson to $\bb$ is the largest, among $\ttH$, the $\ttHbb$ process can be measured with the best statistical precision. However, the final state of the $\ttHbb$ process has an irreducible background from the $\ttbb$ process, which produces a top quark pair in association with a b quark pair. Therefore, understanding the $\ttbb$ process precisely is essential for improving the sensitivity of a search for the $\ttHbb$ process.

For the $\ttbb$ process, the theoretical next-to-leading-order (NLO) calculation was done \cite{bevilacqua2014ratio} in the same phase space where the inclusive cross-sections were measured at $\sqrt{s}$ = 8 TeV in the CMS experiment \cite{cms2014measurement}. This analysis was updated with more data at $\sqrt{s}$ = 13 TeV recently, including inclusive cross-section measurements in the dilepton channel \cite{cms2017measurements}, the lepton+jets channel \cite{sirunyan2020measurement}, and the hadronic channel \cite{sirunyan2020measurement2}.   The inclusive and the differential $\ttbb$ cross-sections were also measured in the ATLAS experiment \cite{miucci2019measurements}. However, these measurements in different channels show consistently that the measured inclusive cross-sections are higher than the theoretical predictions with large uncertainties in both theoretical and experimental results. 

We can provide more information to the theorists to lessen such uncertainties by identifying the origin of the b-jets in the $\ttbb$ process and measuring their respective differential cross-sections. However, this is very challenging in real $\ttbb$ event data because there are no simple rules to distinguish between the b-jets originated from top quark decays and additional b-jets originated from gluon splitting. 

In various high-energy physics problems under similar challenges, machine learning and deep learning techniques have recently been applied to analyze high-dimensional and complex data obtained from the LHC experiments \cite{albertsson2018machine, radovic2018machine, guest2018deep, bourilkov2019machine}. Especially in jet identification problems, of which the goal is to classify the type of jets from data, many deep learning methods have been successful in flavor tagging, jet substructure tagging, and quark/gluon tagging \cite{cogan2015jet, almeida2015playing, de2016jet, louppe2019qcd, larkoski2020jet}. 

A few learning-based attempts have also been made to identify additional b-jets originated from gluon splitting in the $\ttbb$ event.  In the CMS experiment, using early data at $\sqrt{s}$ = 8 TeV, identifying the additional b-jets was attempted for the first time with a boosted decision tree (BDT) in the dilepton channel \cite{cms2016measurement}. Another recent work trained a deep neural network (DNN) as a binary classifier to determine whether a pair of b-jets is a pair of additional b-jets in the lepton+jets channel \cite{choi2020identification}. However, it is difficult to say these approaches are optimal since, during training, they do not exploit the fact that every $\ttbb$ event has only a single pair of additional b-jets. We can take advantage of this particular structure of $\ttbb$ event data by learning to identify only the additional b-jet pair from each event correctly. Such an approach corresponds to directly increasing the matching efficiency, the ratio of successfully identified events to the total number of events. 

In this paper, we apply machine learning techniques to identify two additional b-jets from other b-jets in the $\ttbb$ process. Specifically, we propose methods that can exploit the special structure of $\ttbb$ event data and directly increase matching efficiency. Since the matching efficiency itself is a highly non-smooth objective, we suggest surrogate objective functions suitable for gradient-based optimization. We discuss the difference between directly increasing matching efficiency and the binary classification approach using synthetic data that imitate the $\ttbb$ event data structure. We then show that we can do better in identifying additional b-jets by increasing matching efficiency directly rather than the binary classification accuracy via experiments using simulated $\ttbb$ event data. For the $\ttbb$ event data experiments, we follow the data simulation scheme of \cite{choi2020identification} and consider the lepton+jets channel, which is advantageous for precise measurements due to its large branching fraction, as discussed in \cite{choi2020identification}.

The paper is organized as follows. We define the deep learning-based additional b-jet identification problem in section~\ref{sec:b-jet_id}. We then propose methods to increase matching efficiency directly in section~\ref{sec:max_matching_eff}, discussing its difference to the binary classification approach. Section~\ref{sec:experiments} presents experimental results on simulated $\ttbb$ event data in the lepton+jets channel. 

\section{Deep learning-based additional b-jet identification}
\label{sec:b-jet_id}

\subsection{Problem definition}
\label{sec:problem}
\begin{figure}[tbp]
\centering 
\includegraphics[width=.45\textwidth]{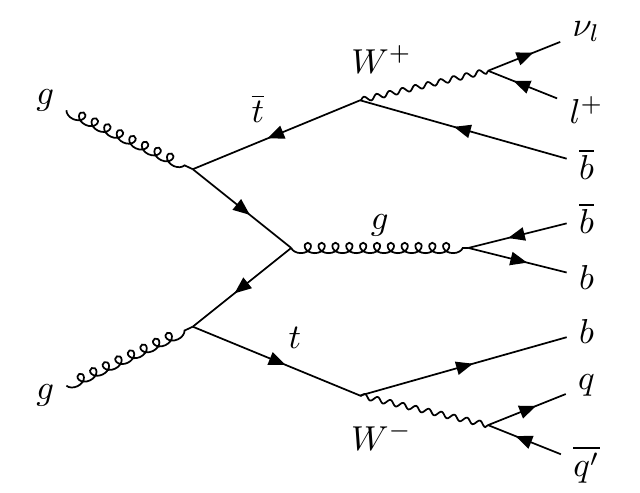}
\caption{\label{fig:ttbb} Feynman diagram of the $\ttbb$ process in the lepton+jets channel.}
\end{figure}

The final state of the $\ttbb$ process contains b-jets from top quark decays and additional b-jets from gluon splitting, as shown in the Feynman diagram in figure~\ref{fig:ttbb}. Our purpose is to precisely predict which of the b-jets correspond to the additional b-jets based on the physical quantities obtained from the final state objects in the process. Due to the high-dimensional nature and complicated stochastic generative processes of the relevant quantities, however, there are no simple rules for identifying additional b-jets. It is also challenging to manually engineer features useful for the task, even if we have some knowledge of the underlying physics model.

Fortunately, when equipped with a large enough data set, discovering useful discriminative features that might even require complex processing on raw data is the point deep learning methods are highly effective. They enable us to learn from data how to transform the input via multiple processing layers of deep neural networks to capture the complicated structure in large data sets and obtain useful features (also called representations) \cite{lecun2015deep}. Therefore we rely on deep learning techniques to achieve our goal; we train deep neural networks that identify additional b-jets under the supervision of a $\ttbb$ event data set in which true additional b-jets are indicated. 

Suppose there are $N$ $\ttbb$ event data. We regard a $\ttbb$ event datum as the collection of every pair of b-jets in the event. Each b-jet pair is represented by an $F$-dimensional vector consisting of physical quantities obtained from its b-jets and other final state objects in the event. (Further details on the selected physical quantities are provided in section~\ref{sec:simulation}.) Denote by $c_i$, $i=1,\ldots,N$ the numbers of b-jet pairs in each event and $M_i \in \mathbb{R}^{c_i \times F}$, $i=1,\ldots,N$ the corresponding matrices consist of all the b-jet pairs (or vectors) in each event. Since the $\ttbb$ process contains only two additional b-jets, there is only a single pair of additional b-jets in each event; let $y_i \in \{1, \ldots, c_i\}$, $i=1,\ldots,N$ be the indices to indicate the pair of additional b-jets. Here the additional b-jet pair becomes the signal to be sought, while the other pairs form the background. 

Our performance criterion is the matching efficiency, the accuracy of identifying additional b-jet pair (or the signal) from each event. Given $N$ $\ttbb$ event data $\{(M_1, y_1), \ldots,$ $(M_N, y_N)\}$, matching efficiency is defined as follows:
\begin{equation}
\label{eq:matching_eff}
\text{Matching efficiency} = \frac{1}{N}\sum_{i=1}^N \delta(y_i, \hat{y}(M_i)), 
\end{equation}
where $\delta(\cdot, \cdot)$ denotes the Kronecker delta function, i.e., $\delta(y, \hat{y}) = 1$ if $y = \hat{y}$ and zero otherwise, and $\hat{y}(M_i)$ denotes the index predicted on the event matrix $M_i$. Increasing matching efficiency becomes our ultimate goal to train the identification models.

\subsection{A previous approach}
\label{sec:previous}
Based on deep learning methods, there was an attempt to identify additional b-jets in the $\ttbb$ events \cite{choi2020identification}. Specifically, they trained a binary classifier to discriminate whether each pair of b-jets is an additional b-jet pair or not. As the training data, they used individual b-jet pairs $x_i \in \mathbb{R}^F$,  $i = 1, \ldots, N_p$ and their corresponding labels $\xi_i \in \{0, 1\}$, $i = 1, \ldots, N_p$ separated from all the training event data $\{(M_1, y_1), \ldots, (M_N, y_N)\}$ (hence the number of training data $N_p = \sum_{i=1}^N c_i$). Here the labels are 1 for additional b-jet pairs and 0 otherwise. Note that, when collecting pairs from a single event, there is only one pair with the label of 1 since every $\ttbb$ event data has a single additional b-jet pair.

The binary classifier is modeled as a deep feedforward neural network $f: \mathbb{R}^F \rightarrow [0,1]$ as detailed in appendix~\ref{appendix:MLP}. Given an input b-jet pair, it returns the value between zero and one, which can be conceptually interpreted as the probability for the input pair to be label 1. The model parameters are trained to minimize the binary cross-entropy loss ($L_{BCE}$) defined as follows:\footnote{To be precise, the objective function in eq.~\eqref{eq:BCE} is the sample average of the losses $l(\xi_i, f(x_i)) = - \xi_i\log f(x_i) - (1-\xi_i)\log (1-f(x_i))$,  $i = 1, \ldots, N_p$, but we denote such an average by `loss' for simplicity.}
\begin{equation}
\label{eq:BCE}
L_{BCE} = -\frac{1}{N_p} \sum_{i=1}^{N_p} \xi_i \log f(x_i) + (1-\xi_i) \log (1-f(x_i)).    
\end{equation}
When a new event comes in after training the model, they select the pair with the highest model output value as the pair of additional b-jets. 

Even though the suggested method showed better performance than another method using a physics-based feature, it is hard to say that this method is optimal for increasing the matching efficiency. This is because the objective function in eq.~\eqref{eq:BCE} is designed to increase the binary classification accuracy for all b-jet pairs, which does not precisely match our goal. Furthermore, the method overlooked the particular structure of the input during training that there exists only one b-jet pair with the label of 1 in each event. Leveraging this input structure, we can directly increase matching efficiency by identifying only the additional b-jet pair in each event correctly. We now propose methods to implement this idea and also compare them with the binary classification approach.

\section{Directly maximizing matching efficiency}
\label{sec:max_matching_eff}

\subsection{Prediction model}
\label{sec:pred_model}

We first propose the form of our prediction model to be used in maximizing matching efficiency directly. Given event matrices $M_i \in \mathbb{R}^{c_i \times F}$, $i=1,\ldots,N$ as the input, our model $f: \mathbb{R}^{c_i \times F} \rightarrow [0,1]^{c_i}$ is set to\footnote{Hence our model should deal with varying sizes of input and output according to $c_i$, $i=1,\ldots,N$.} 
\begin{equation}
\label{eq:model}
f = (f_1, \ldots, f_{c_i}),
\end{equation}
where the output of each $f_j: \mathbb{R}^{c_i \times F} \rightarrow [0,1]$ can be interpreted as the probability of the $j$-th b-jet pair (or the $j$-th row of $M_i$) being the pair of additional b-jets for $j=1,\ldots,c_i$. Here the output elements should sum to one, i.e., $\sum_{j=1}^{c_i} f_j(M_i) = 1$, so that the output $f(M_i)$ can be a proper distribution for $i=1,\ldots,N$. 

By defining the output as such a probability distribution, this model inherently reflects the presence of one additional b-jet pair in every $\ttbb$ event. Moreover, the additional b-jet pair is straightforwardly predicted to be the pair that returns the highest probability, i.e.,
\begin{equation}
\label{eq:predict}
\hat{y}(M_i) = \argmax_{j \in \{1, \ldots, c_i\}} f_j(M_i), \quad i=1,\ldots,N.    
\end{equation}
We model such an $f$ using deep neural networks as detailed in appendix~\ref{appendix:details_pred}.

\subsection{Surrogate loss functions}
When training the prediction model $f$, it is not easy to maximize the matching efficiency in eq.~\eqref{eq:matching_eff} itself since it is a highly non-smooth objective, i.e., not differentiable in many different regions of the model parameter space, and even returns zero gradients when differentiable. Therefore, we propose appropriate relaxations or surrogates of eq.~\eqref{eq:matching_eff} suitable for usual gradient-based optimization while still inducing the model to increase matching efficiency directly.

In terms of our model presented in the previous section, observe that the $\delta(y_i, \hat{y}(M_i))$ value in eq.~\eqref{eq:matching_eff} becomes one if $f_{y_i}(M_i)$ is the largest among $\{f_1(M_i), \ldots, f_{c_i}(M_i)\}$ and zero otherwise according to eq.~\eqref{eq:predict}. Hence we can choose to maximize $f_{y_i}(M_i)$ instead of $\delta(y_i, \hat{y}(M_i))$ to make the $\delta(y_i, \hat{y}(M_i))$ value be one and consequently increase the matching efficiency. Since the usual machine learning problems are formulated as `minimizing losses,' we attach the negative sign and define our first surrogate loss function (denoted $L_1$) as follows:
\begin{equation}
\label{eq:L1}
L_1 = -\frac{1}{N}\sum_{i=1}^N f_{y_i}(M_i).
\end{equation}

Similarly, we can maximize model outputs for the additional b-jet pair while minimizing those for the other pairs simultaneously. The corresponding loss function (denoted $L_2$) is defined as follows:
\begin{equation}
\label{eq:L2}
L_2 = -\frac{1}{N}\sum_{i=1}^N f_{y_i}(M_i) - \sum_{j\neq y_i} f_j(M_i).
\end{equation}

From a slightly different perspective, it is possible to deem an event matrix's row indices as distinctive categories to which the additional b-jet pair belongs. The problem can then be viewed as a multi-class classification problem to predict the class, i.e., the additional b-jet pair index, for each event matrix; corresponding classification accuracy becomes exactly the same as the matching efficiency in eq.~\eqref{eq:matching_eff}. Hence, the widely used (categorical) cross-entropy loss (denoted $L_3$) for these classification problems can also serve as our surrogate loss:
\begin{equation}
\label{eq:L3}
L_{3} = -\frac{1}{N}\sum_{i=1}^N \log f_{y_i}(M_i).
\end{equation}

Based on this interpretation, though not used as often as the cross-entropy loss in classification problems, we can also consider minimizing the mean squared error loss (denoted $L_4$) defined as:
\begin{equation}
\label{eq:L4}
L_{4} = \frac{1}{N} \sum_{i=1}^N \Vert f(M_i) - \text{one}\_ \text{hot}(y_i)\Vert^2,
\end{equation}
where $\text{one}\_\text{hot}(y_i) = [0, \ldots, 1, \dots, 0] \in \mathbb{R}^{c_i}$ is the one-hot encoding whose $y_i$-th element is one.

\subsection{Comparison to binary classification via synthetic data experiments}

\begin{figure}[tbp]
     \centering
     \begin{subfigure}[b]{0.23\textwidth}
         \centering
         \includegraphics[width=\textwidth]{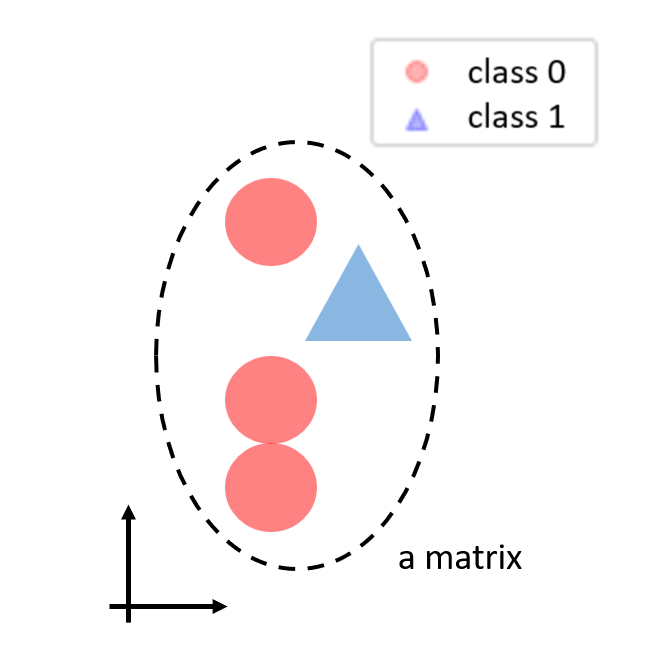}
         \caption{An example matrix}
         \label{fig:syn_event}
     \end{subfigure}
     \hfill
     \begin{subfigure}[b]{0.35\textwidth}
         \centering
         \includegraphics[width=\textwidth]{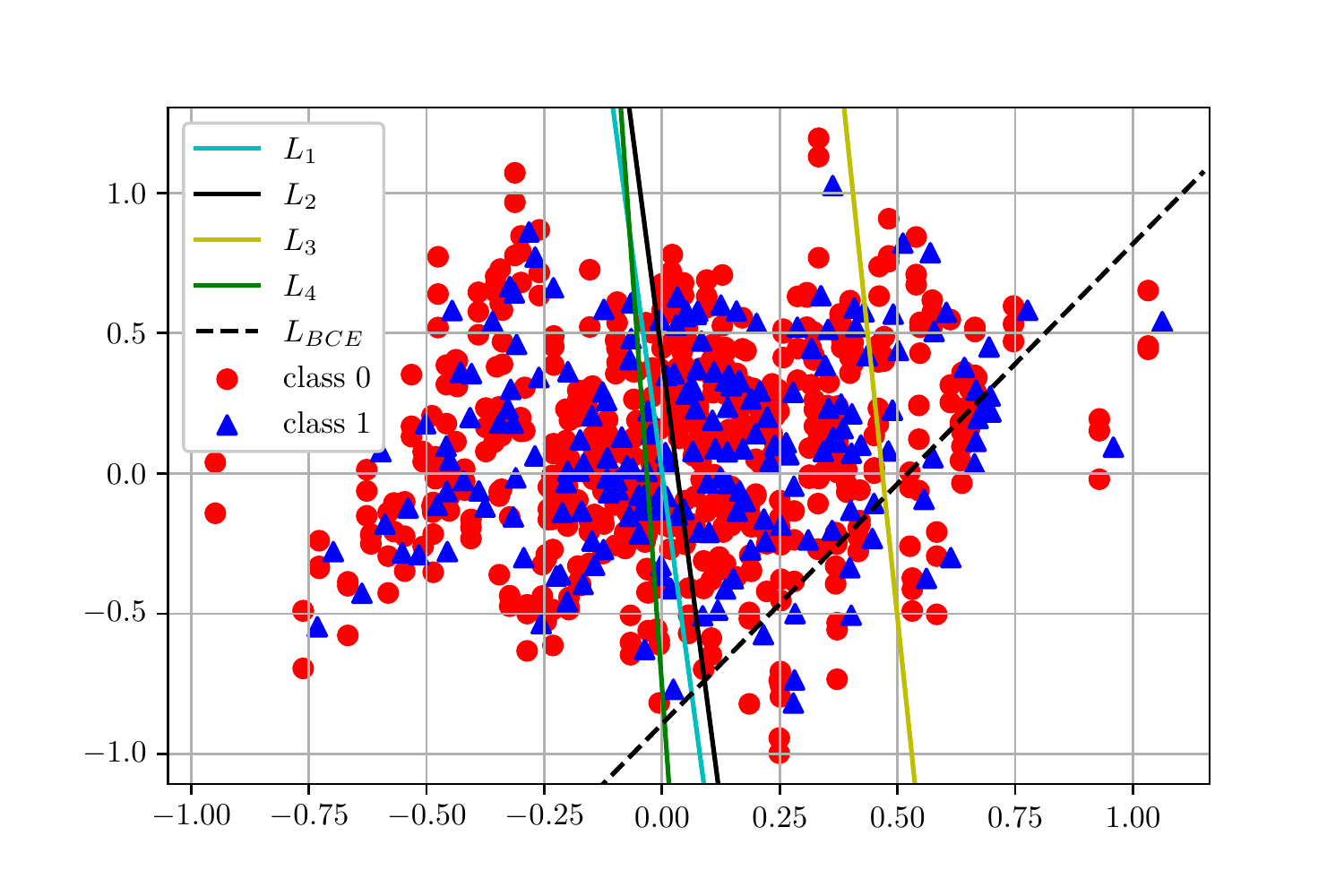}
         \caption{Trained models}
         \label{fig:syn_scatter}
     \end{subfigure}
     \hfill
     \begin{subfigure}[b]{0.35\textwidth}
         \centering
         \includegraphics[width=\textwidth]{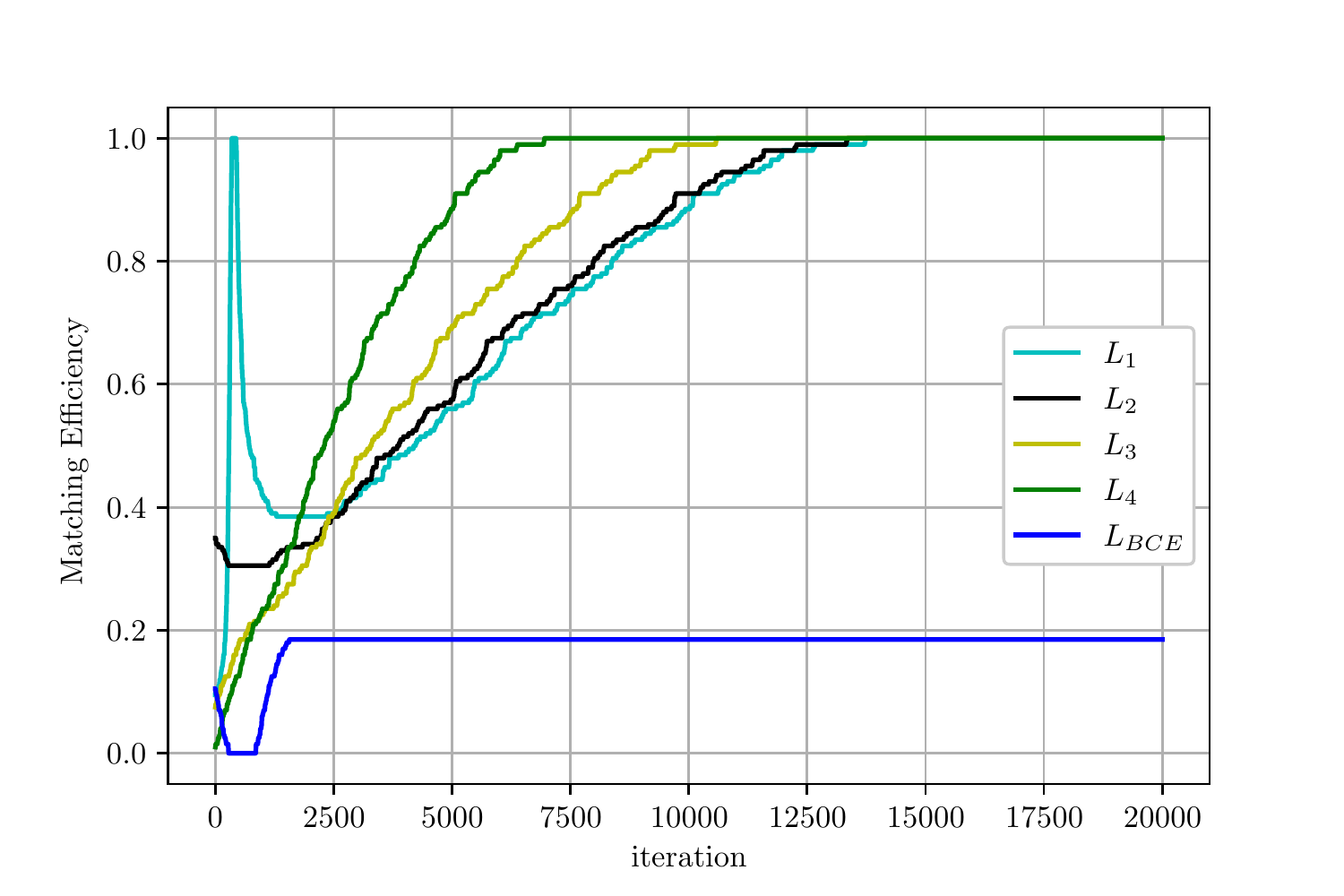}
         \caption{Matching efficiency}
         \label{fig:syn_match_eff}
     \end{subfigure}
        \caption{Experimental results from synthetic data. An example matrix is illustrated (left). The level sets of the trained models are drawn over the data scatter plot (middle). Matching efficiency (evaluated on a test data set) for each of the loss functions are shown (right).}
        \label{fig:syn_results}
\end{figure}

We now conduct an experiment using synthetic data to provide an insight into how minimizing the proposed loss functions can differ from the binary classification approach. For this purpose, we generate two-dimensional data points and construct matrices from them to imitate the constraint of $\ttbb$ event data, which have only a single row vector of label 1 in each matrix. Specifically, each matrix consists of a vector in class 1 sampled from the two-dimensional normal distribution and three vectors in class 0 obtained by translating the class 1 vector to the left slightly and injecting noise along the vertical axis (as shown in figure~\ref{fig:syn_event}). This setup will make the binary classification approach suffer from a significant overlap between data from classes 0 and 1 since, during training, all vectors are given mixed regardless of their originated matrices, as shown in figure~\ref{fig:syn_scatter}.

We train simple linear models to identify class 1 data by minimizing the suggested loss functions in eqs.~\eqref{eq:L1}, \eqref{eq:L2}, \eqref{eq:L3}, \eqref{eq:L4}, and the binary cross-entropy loss in eq.~\eqref{eq:BCE}. (Further details of the experiments are explained in appendix~\ref{appendix:details_syn}.) For each of the trained models, we plot their decision boundary or the level sets, i.e., the set of input points with identical output values, in figure~\ref{fig:syn_scatter}. Considering the structure of the matrices, the models should return larger outputs for the points with larger horizontal axis values for successful identification; hence the vertical decision boundary or level sets are desired. However, the binary classification approach gives a somewhat arbitrary decision boundary due to the significant overlap between classes. On the other hand, observe that the level sets of the models obtained from minimizing eqs.~\eqref{eq:L1}, \eqref{eq:L2}, \eqref{eq:L3}, \eqref{eq:L4} are almost vertical as desired. Thanks to the loss functions formulated to exploit the special structure of the data, these models can learn how to discriminate the data inside each matrix.

The matching efficiency of each model is evaluated on a test data set sampled in the same way as the training data set. According to figure~\ref{fig:syn_match_eff}, the models trained using the proposed loss functions obtain matching efficiency of one, i.e., identify the class 1 data perfectly for all matrices, while that from the binary classification approach shows a much lower value. Although the setup is much simpler than that of the $\ttbb$ event data, this experiment effectively verifies that minimizing the suggested loss functions can increase matching efficiency better than the binary classification approach.

\section{Experiments on simulated \texorpdfstring{\boldmath{$\ttbb$}}{ttbb} event data}
\label{sec:experiments}

\subsection{Simulated data set}
\label{sec:simulation}
We now examine if the proposed methods can achieve better matching efficiency than the binary classification approach on simulated $\ttbb$ event data. Here we follow the overall simulation scheme of \cite{choi2020identification} to generate data. The simulated $\ttbb$ events in pp collisions are produced at a center-of-mass energy of 13 TeV. We generated 16 million events for the $\ttbb$ samples by using the $\textsc{MadGraph5aMC@NLO}$ program (v2.6.6) \cite{alwall2014automated} at the leading order, and these events are further interfaced to $\textsc{Pythia}$ (v8.240) \cite{sjostrand2015introduction} for the hadronization. A W boson decays through $\textsc{MadSpin}$ \cite{artoisenet2013automatic}, and the events are generated in a 4-flavor scheme, where the b quark has mass.

The generated events are processed using the detector simulation with the DELPHES package (v3.4.1) \cite{de2014delphes} for the CMS detector. The physics objects used in this analysis are reconstructed based on the particle-flow algorithm \cite{sirunyan2017particle} implemented in the DELPHES framework. In the DELPHES fast simulation, the final momenta of all the physics objects, such as electrons, muons, and jets, are smeared as a function of both the transverse momentum $p_T$ and the pseudorapidity $\eta$ so that they can represent the detector effects. The efficiencies of identifying the electrons, muons, and jets are parameterized as functions of $p_T$ and $\eta$ based on information from the measurements made by using the CMS data \cite{de2014delphes}. The particle-flow jets used in this analysis are reconstructed using the anti-$k_T$ algorithm \cite{cacciari2008anti} to cluster the particle-flow tracks and particle-flow towers.

The b-tagging efficiency is around 50$\%$ at the tight-working point of the deep combined secondary vertex (DeepCSV) algorithm \cite{guest2016jet}, which shows the best performance in the CMS measurement \cite{sirunyan2018identification}. The corresponding fake b-tagging rates from the c-flavor and the light flavor jets are set to around 2.6$\%$ and 0.1$\%$, respectively. Once events are produced, the $\ttbb$ process is defined based on the particle-level jets obtained by clustering all final-state particles at the generator level. A jet is considered as an additional b-jet if the jet is matched to the last b quark that is not from a top quark within $\Delta R(j, q) = \sqrt{\Delta \eta(j, q)^2 + \Delta\phi(j, q)^2} < 0.5$, where $j$ denotes jets at the generator level, and $q$ denotes the last b quark. The additional b-jets are required to be within the experimentally accessible kinematic region of $p_T$ > 20 GeV and $|\eta| < 2.5$. Under the condition that there should exist at least two additional b-jets to be a valid $\ttbb$ event, 18$\%$ of the generated samples remain.

We applied the following event selection to remove the main backgrounds from the multi-jet events and W+jet events. At the reconstruction level of the lepton+jets channel, the event must have exclusively one lepton with $p_T$ > 30 GeV and $\eta < 2.4$. According to this condition, 5.6$\%$ of the generated events survive. Jets are selected with a threshold of $p_T$ > 30 GeV and $|\eta| < 2.5$. The $\ttbb$ event has the final state of four b-jets (including the two additional b-jets) and two jets from one of the two W bosons in top quark decays. However, the detector acceptance and the efficiencies of the b-jet tagging algorithms are not 100$\%$; hence some of the $\ttbb$ events have fewer jets at the reconstruction level. Therefore, we discarded the events containing b-jets fewer than three or jets fewer than six in our experiments, resulting in 45,501 matchable events that include two additional b-jets.

After the event selection, we construct data for each $\ttbb$ event by gathering every pair of b-jets in the event, as explained in section~\ref{sec:problem}. Specifically, the variables representing a pair of b-jets are selected considering all possible combinations of the four-vectors of the final state objects, such as selected two b-tagged jets, a lepton, a reconstructed hadronic W boson, and missing transverse energy (MET), which are considered low-level features. We consider the total of 78 variables as listed in section~V of \cite{choi2020identification}.

\subsection{Experimental details}

We train our model defined in section~\ref{sec:pred_model} on simulated $\ttbb$ event data by minimizing the proposed loss functions. When constructing a deep neural network for eq.~\eqref{eq:model} (or $g$ in eq.~\eqref{eq:simplification} to be specific), we utilize the dropout layer \cite{srivastava2014dropout} to prevent overfitting on training data, and the batch normalization layer \cite{ioffe2015batch} to increase the training speed. Adam optimizer \cite{kingma2014adam} is applied to train the models with a batch size of 2,048 and a learning rate of 1e-3. We conducted all the experiments using Keras \cite{chollet2015keras}.

The data set is split into training, validation, and test sets with respective sizes as 35,398/4,358/5,745.
We further divide the training set into the primary and additional sets with respective sizes as  17,988/17,410 to investigate the effect of increasing the training data size;  we denote by $\mathcal{D}_1$ the primary training and validation sets and $\mathcal{D}_2$ the additional training set.

We use the validation set to determine hyperparameters of the deep neural network, such as the number of hidden layers $L$ and the dimension of hidden variables $d$; we consider $L \in \{2, 4, 6, 8, 10\}$ and $d \in \{25, 50, 100, 200\}$ in the search.  Early stopping is also used to avoid overfitting and determine the appropriate number of epochs for training using an algorithm provided in \cite{goodfellow2016deep}. 

After the hyperparameter tuning, we train models by minimizing the proposed loss functions according to two different training data configurations of using (i) the primary training and validation sets ($\mathcal{D}_1$), and (ii) all the training and validation sets ($\mathcal{D}_1 + \mathcal{D}_2$). Models trained to minimize the binary cross-entropy loss are also considered for comparison purposes. The final performance on the test set is then averaged over twenty runs (with different initializations for the model parameters). For the performance metric, we consider the matching efficiency defined in eq.~\eqref{eq:matching_eff} and the area under the ROC curve (AUC), a widely used metric for the binary classification.

\subsection{Experimental results}

The matching efficiency and the AUC score of the trained models are reported in table~\ref{tab:ttbb_result}. We can observe that minimizing binary cross-entropy loss yields the best AUC score, but this does not necessarily lead to the best matching efficiency. Compared to the binary classification approach, minimizing the proposed loss functions shows performance gains ranging between 0.5--1.7 percent points in average matching efficiency. Interestingly, under the training data configuration of $\mathcal{D}_1$, such performance gains (1.2--1.7 percent points) are even larger than those from using additional training data in the binary classification approach (0.9 percent points). Among the proposed loss functions, the cross-entropy loss ($L_3$) shows the best performance with average matching efficiency of 0.625 under the training data configuration of $\mathcal{D}_1$ + $\mathcal{D}_2$. From these results, we can conclude that simply changing the loss functions to directly increase matching efficiency (especially using $L_3$) shows a definite performance improvement in identifying additional b-jets in the $\ttbb$ event data.

\begin{table}[tbp]
\begin{center}
\resizebox{0.9\textwidth}{!}{%
\begin{tabular}{c|cc|cc}
    \hline
    \multirow{2}{*}{Loss functions} & \multicolumn{2}{c|}{$\mathcal{D}_1$} & \multicolumn{2}{c}{$\mathcal{D}_1+\mathcal{D}_2$} \\
    \cline{2-5}
     & Mat. eff. & AUC & Mat. eff. & AUC \\
     \hline\hline
    $L_{\text{BCE}}$ & 0.602 (4.31e-3) & 0.756 (2.10e-3) & 0.611 (3.12e-3) & \textbf{0.764} (1.39e-3) \\
    \hline
    $L_{1}$          & 0.614 (2.65e-3) & 0.727 (2.52e-3) & 0.616 (2.09e-3) & 0.729 (3.44e-3) \\
    \hline
    $L_{2}$          & 0.615 (2.18e-3) & 0.732 (2.51e-3) & 0.616 (1.60e-3) & 0.731 (2.30e-3) \\
    \hline
    $L_{3}$          & 0.619 (2.68e-3) & 0.739 (6.04e-3) & \textbf{0.625} (1.86e-3) & 0.741 (2.33e-3) \\
    \hline
    $L_{4}$          & 0.616 (2.33e-3) & 0.734 (2.10e-3) & 0.617 (2.55e-3) & 0.733 (3.96e-3) \\
    \hline
\end{tabular}
}
\end{center}
\caption{Average matching efficiency (Mat. eff.) and AUC on the test set from different training configurations with standard deviations in the parentheses.}
\label{tab:ttbb_result}
\end{table}

\begin{figure}[tbp]
     \centering
     \begin{subfigure}[b]{0.45\textwidth}
         \centering
         \includegraphics[width=\textwidth]{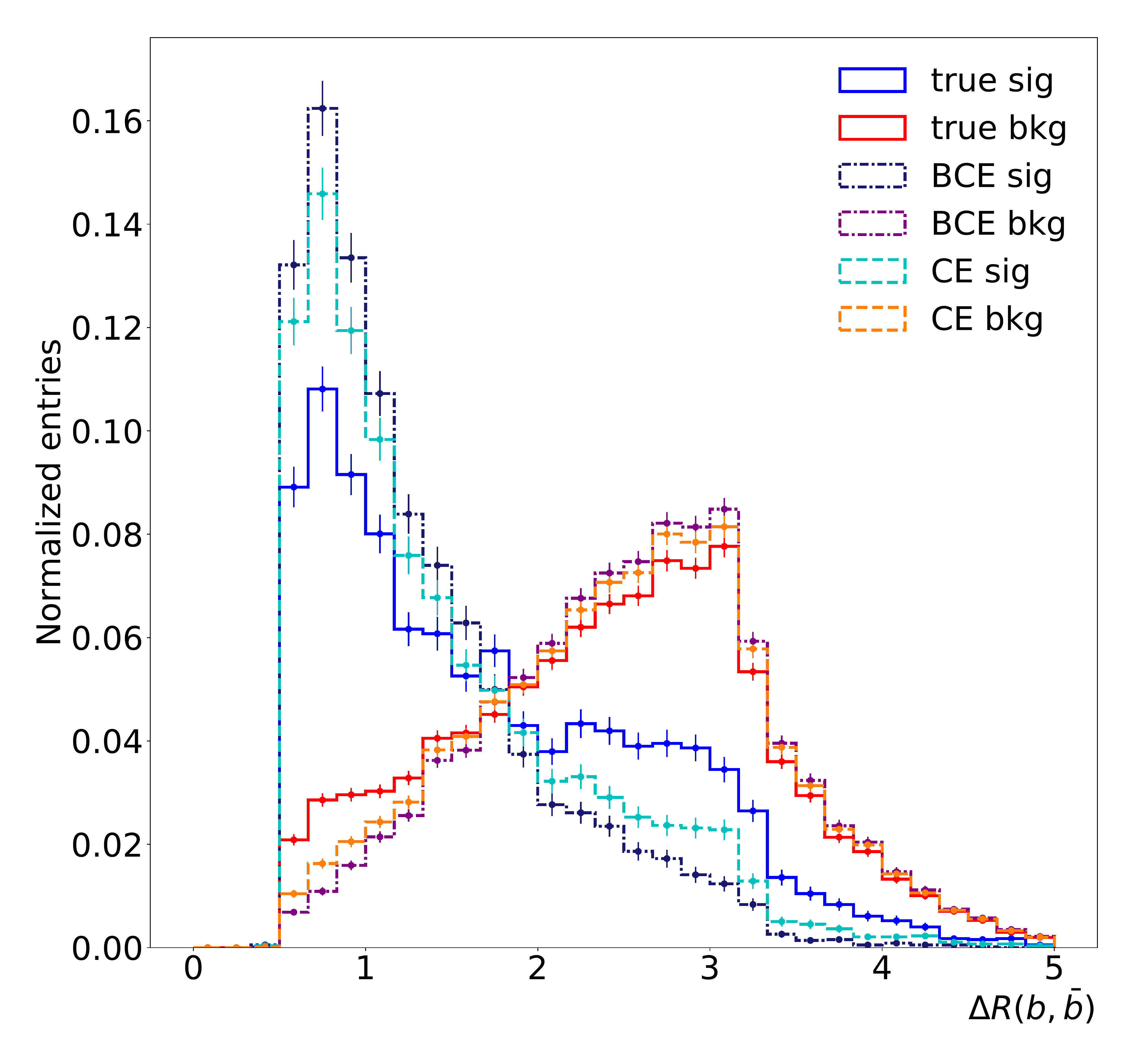}
         \caption{$\Delta R$}
         \label{fig:ttbb_bbdR}
     \end{subfigure}
     \hfill
     \begin{subfigure}[b]{0.45\textwidth}
         \centering
         \includegraphics[width=\textwidth]{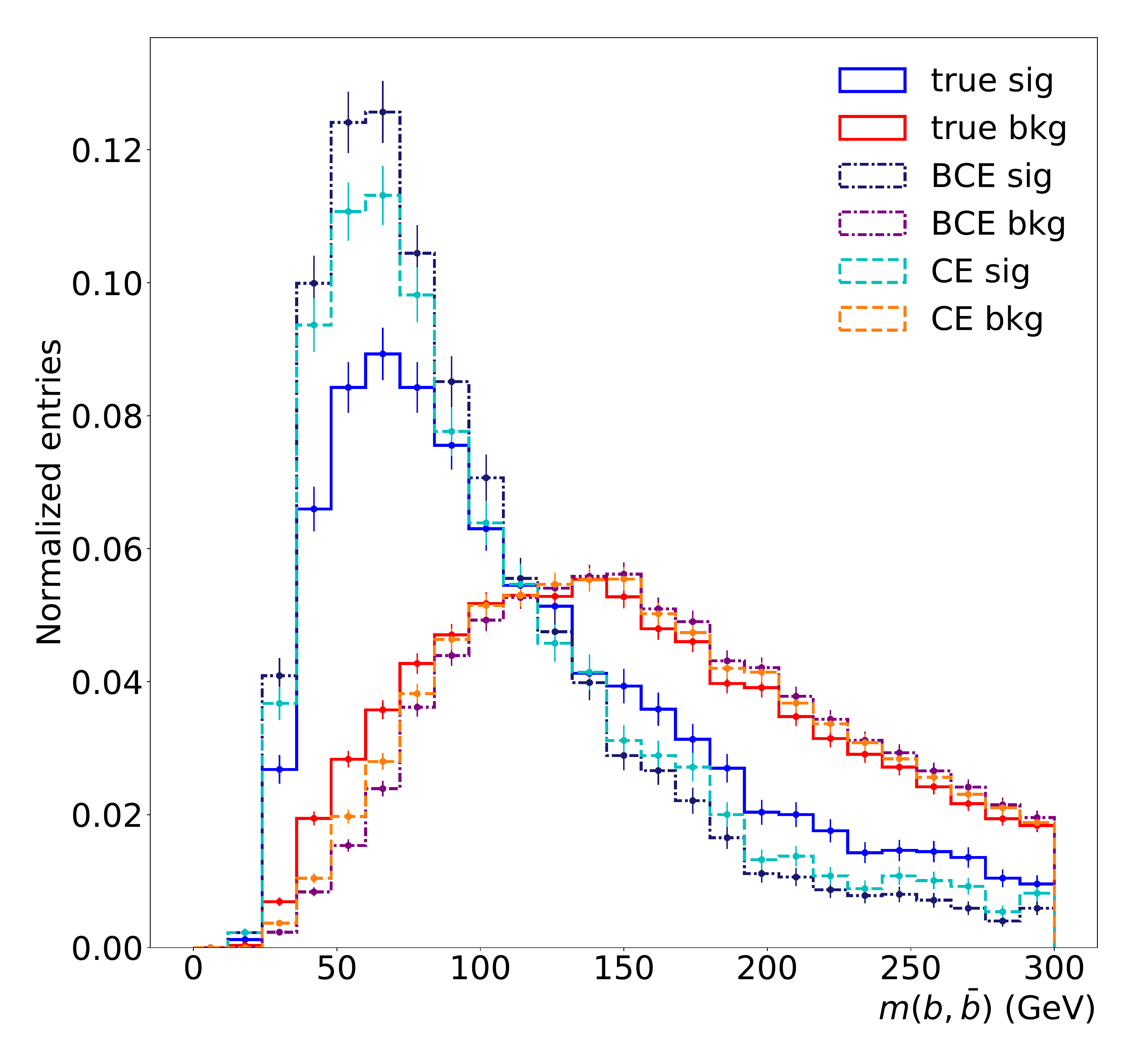}
         \caption{Invariant mass}
         \label{fig:ttbb_bbMass}
     \end{subfigure}
        \caption{Histograms of $\Delta R$ (angular distance) and invariant mass of b-jet pairs in the test set. The additional b-jet pairs are denoted the signal (sig), and the other b-jet pairs are denoted the background (bkg). Additional b-jet pairs are identified according to three different criteria (true: the true label, BCE: the prediction from the binary classification approach, CE: the prediction from our method using $L_3$, i.e., the cross-entropy loss).}
        \label{fig:ttbb_analysis}
\end{figure}

We can also examine the identification performance qualitatively by observing distributions of physical quantities obtained from b-jet pairs. In figure~\ref{fig:ttbb_analysis}, we obtain histograms of both $\Delta R$ (the angular difference) and the invariant mass for the additional b-jet pairs and the other pairs identified from the test set. Here we consider three different identification results based on (i) the true label, (ii) the prediction from the binary classification approach, and (iii) the prediction from our method using $L_3$ (the cross-entropy loss). Compared to the binary classification approach, our method using $L_3$ better captures the histograms obtained from the true label, demonstrating increased identification performance. Since observable improvements in the histograms occur relatively uniformly across the entire bins, our identification performance improvements can be considered plausible from a physical viewpoint as well.

\section{Conclusions}
In measuring the differential cross-sections of the $\ttbb$ process from the properties of its b-jets, it is crucial to identify additional b-jets originated from gluon splitting correctly. In this paper, we have proposed different loss functions to directly increase matching efficiency, the accuracy of identifying additional b-jets. Unlike the previous deep learning-based binary classification approach in \cite{choi2020identification}, these loss functions are designed to fully exploit the special structure of the $\ttbb$ event data, as discussed via a simple synthetic data experiment. Using the simulated $\ttbb$ event data in the lepton+jets channel from pp collision at $\sqrt{s}$ = 13 TeV, we have also verified that directly maximizing the matching efficiency via our method shows better performance in identifying additional b-jets than the previous approach. 

In the future, our simulation can be extended in line with the settings of Run-3 at the LHC or the High Luminosity LHC (HL-LHC) to verify our method's applicability. To further improve the identification performance, we can consider utilizing more sophisticated high-level features such as the b-tag discriminator. Modifying the neural network architectures to obtain better discriminative features, e.g., by mixing features from different b-jet pairs, would be another intriguing option. These future works would ultimately lead to a more precise understanding of the $\ttbb$ process and help search the $\ttHbb$ process to study the properties of the Higgs boson.

\appendix
\section{The deep feedforward neural network}
\label{appendix:MLP}
A deep feedforward neural network $f: \mathbb{R}^m \rightarrow \mathbb{R}^n$ with $L$ hidden layers can be modeled as follows:
\begin{equation}
\label{eq:bc_model}
\begin{split}
h_{i} & = \sigma_{i}(W_i h_{i-1} + b_i), \qquad i = 1, \ldots, L, \\
f(x) & = \sigma(W_{L+1} h_{L} + b_{L+1}),  
\end{split}
\end{equation}
where $h_i \in \mathbb{R}^d$, $i=1,\ldots,L$ denote the hidden variables, $h_0$ is set to be the input $x \in \mathbb{R}^m$, $\sigma_i(\cdot)$, $i=1,\ldots,L$ denote nonlinear activation functions, and $W_i, b_i$ for $i = 1, \ldots, L+1$ respectively denote the matrix and vector parameters with sizes defined accordingly as above. In the case of binary classification, the output dimension $n$ is set to one and the activation function $\sigma(\cdot)$ for the output is usually chosen to be the sigmoid function $\sigma(s) = \frac{1}{1+\exp(-s)}$. For general $n$-dimensional vector outputs, $\sigma(\cdot)$ is set to be an identity. 


\section{Details for the prediction model}
\label{appendix:details_pred}
The constraint $\sum_{j=1}^{c_i} f_j(M_i) = 1$ on the prediction model in eq.~\eqref{eq:model} is usually realized by using the softmax function on some activation value $(g_1(M_i), \ldots, g_{c_i}(M_i)) \in \mathbb{R}^{c_i}$ as
\begin{equation}
\label{eq:softmax}
f_j(M_i) = \frac{\exp(g_j(M_i))}{\sum_{k=1}^{c_i}\exp(g_k(M_i))}, \quad j=1,\ldots, c_i,    
\end{equation}
where $g_j: \mathbb{R}^{c_i \times F} \rightarrow \mathbb{R}$ denotes a function to return the $j$-th activation value. In modeling $g_j$, $j=1,\ldots,c_i$, we make another simplification to deal with different $c_i$s (or the numbers of b-jet pairs) for each datum as follows:
\begin{equation}
\label{eq:simplification}
    g_j(M_i) = g(M_{i,j:}), \quad j=1,\ldots,c_i,
\end{equation}
where $M_{i,j:} \in \mathbb{R}^F$ denotes the $j$-th row of $M_i\in \mathbb{R}^{c_i \times F}$, and $g: \mathbb{R}^{F} \rightarrow \mathbb{R}$ can be modeled as a deep neural network as explained in appendix~\ref{appendix:MLP}. Note that this model applies identical operations to each b-jet pair (or row vector) and can have similar modeling complexity to the binary classifier explained in section~\ref{sec:previous}.

\section{Details for synthetic data experiments}
\label{appendix:details_syn}
 To generate synthetic data, we first sample $N$ data in class 1 from the two-dimensional normal distribution, i.e., $x_{1i} \sim N(0,I)$, $i=1,\ldots,N$ with $I\in \mathbb{R}^{2\times 2}$ as the identity matrix. We draw three data points in class 0 by translating each class 1 datum along the horizontal axis for a constant and then inject the Gaussian noise along the vertical axis, in specific, $x_{0i,j} = x_{1i} + (-0.03, \epsilon_j) \in \mathbb{R}^2$ with $\epsilon_j \sim N(-0.01,0.1)$ for $i = 1, \ldots, N$, $j = 1,2,3$. We then randomly stack the vectors $x_{0i,1}, x_{0i,2}, x_{0i,3}, x_{1i}\in\mathbb{R}^2$ to construct each matrix $M_i \in \mathbb{R}^{4\times 2}$ for $i = 1, \ldots, N$.
 
For the approaches directly maximizing matching efficiency using the surrogate losses in eqs.~\eqref{eq:L1}, \eqref{eq:L2}, \eqref{eq:L3}, \eqref{eq:L4}, the function $g: \mathbb{R}^{2} \rightarrow \mathbb{R}$ in eq.~\eqref{eq:simplification} is defined as $g(x) = x W + b$, where $W \in \mathbb{R}^{2\times 1}, b \in \mathbb{R}$ are the model parameters. For the binary classification approach that minimizes eq.~\eqref{eq:BCE}, we consider the model $f: \mathbb{R}^2 \rightarrow [0,1]$ defined as $f(x) = \sigma(xW + b)$, where $W, b$ are the model parameters of identical size as the above, and $\sigma(\cdot)$ denotes the sigmoid function.

\acknowledgments
JC was supported by IITP Artificial Intelligence Graduate School Program for Hanyang University funded by MSIT (Grant No. 2020-0-01373).
NYK was partly supported by NRF/MSIT (No. 2018R1A5A7059549),  IITP Artificial Intelligence Graduate School Program for Hanyang University (2020-0-01373), and Hanyang University (HY-2019).
KTJ was supported by the Basic Science Research Program through the National Research Foundation of Korea (NRF) funded by the Ministry of Education (Grant No. NRF-2020R1A2C2005228).



\bibliographystyle{JHEP}
\bibliography{paper}







\end{document}